\documentstyle[12pt]{article}
\newlength{\mytopmargin}
\newlength{\myleftmargin}
\begin{document}
\noindent
\begin{center}{  \Large\bf
 EXACT AND ASYMPTOTIC FORMULAS FOR\\[2mm] OVERDAMPED
  BROWNIAN DYNAMICS}
\end{center}
\vspace{5mm}

 \noindent
 \begin{center}
 P.J.~Forrester\footnote{Permanent address:  Department of Mathematics, 
University of Melbourne, 
  Parkville, Victoria 3052, Australia. email: matpjf@maths.mu.oz.au} 
and 
B.~Jancovici\footnote{email: janco@stat.th.u-psud.fr} \\
Laboratoire de Physique Th\'eorique et Hautes Energies, \\
Universit\'e de Paris-Sud, 91405 Orsay Cedex, France\footnote{Laboratoire
associ\'e au Centre National de la Recherche Scientifique. URA D0063.}
\end{center}
 \vspace{.5cm}

\begin{quote}
Exact and asymptotic formulas relating to dynamical correlations for overdamped
Brownian motion are obtained. These formulas include a generalization of the
$f$-sum rule from the theory of quantum fluids, a formula relating the static
current-current correlation to the static density-density correlation, and an
asymptotic formula for the small-$k$ behaviour of the dynamical structure
factor. Known exact evaluations of the dynamical density-density correlation
in some special models are used to illustrate and test the formulas.
\end{quote}

\section{Introduction}
A number of recent papers [1-4] have considered various aspects of the density
and current correlations in the Dyson Brownian motion model [5]. This model
refers to the overdamped Brownian dynamics of the one-dimensional
one-component log-gas. When the initial state is given by the
equilibrium state, the model is
 equivalent [1] to the ground state dynamics of the 
quantum many body system with
$1/r^2$ two-body interactions (Calogero-Sutherland model). One motivating factor
for the current interest is that the Dyson Brownian motion model specifies the
eigenvalue probability density function (p.d.f.) for certain ensembles of
parameter dependent random matrices (see e.g.~[6]). These random matrix
ensembles are used to study the parametric dependence (typically as a function
of magnetic field strength) of quantum energy spectra 
and fluctation phenomena in
quantum transport problems [7,8]. Another motivating factor is that there are
now a number of exact results available for the dynamical density-density
correlation [4,7-9].

In this paper we seek to extend the considerations of earlier works by
considering properties of dynamical correlations for overdamped Brownian
motion described by the general Fokker-Planck equation
$$
\gamma {\partial p \over \partial \tau} ={\cal L}p \qquad \mbox{where} \quad
{\cal L} = \sum_{j=1}^N {\partial  \over \partial x_j}
\left (   {\partial W \over \partial x_j}
+\beta^{-1} {\partial  \over \partial x_j} \right ).
\eqno (1.1)
$$
The Dyson  Brownian motion model is specified by this equation with  the
particular potential
$$
W = {1 \over 2}
\sum_{j=1}^N x_j^2 - \sum_{1 \le j < k \le N} \log | x_k - x_j|,
\eqno (1.2a)
$$
or its periodic version
$$
W =  - \sum_{1 \le j < k \le N}
\log |e^{2 \pi i x_k/L} - e^{2 \pi i x_j/L}|.
\eqno (1.2b)
$$

We begin Section 2 by reviewing properties of (1.1). We then clarify the
meaning of the current (notice that (1.1) is independent of the velocities).
Next a generalization of the $f$-sum rule for quantum fluids [10] is given,
as is a  generalization of the formula of Taniguchi et al.~[3] relating
the initial density-density and current-current correlations. In the final
two subsections of Section 2 we consider the hydrodynamic approximation to the
microscopic density fluctuations, and derive from it the
small-$k$ asymptotics of the density-density and current-current
correlations, which in turn imply certain sum rules and further
asymptotic formulas. The analytic formulas of Section 2 are 
illustrated and tested on some exact results for the density-density correlation
in the Dyson  Brownian motion model (i.e.~(1.1) with potential (1.2)), and
for (1.1) with potentials closely related to (1.2). Our results are
summarized in Section 4.

\section{The density and current correlations}
The Fokker-Planck equation (1.1) relates to the overdamped Brownian motion
dynamics of a classical gas with potential energy $W$ in contact with a heat
bath at inverse temperature $\beta$. The function $p$ is the p.d.f.~for the
event that the particles are at positions $x_1,\dots,x_N$ after time
$\tau$. As written in (1.1), the particles are assumed to be confined to a
line, however the same equation applies in higher space dimensions if we
simply replace $x_j$ by $x_j^{(\alpha)}$ (the components of the position
coordinate $\vec{x}_j$) and assume summation over $\alpha$.

The dynamics specified by (1.1) is known [11] to be equivalent to the
dynamics specified by the coupled Langevin equations
$$
\gamma {d x_j(\tau) \over d \tau} = - {\partial W \over \partial x_j}
+ {\cal F}_j(\tau) \qquad (j=1,\dots, N)
\eqno (2.1a)
$$
where the random force ${\cal F}_j(\tau)$ is a Gaussian random variable
with zero mean and variance given by
$$
\overline{{\cal F}_i(\tau) {\cal F}_j(\tau')} = {2 \gamma \over \beta}
\delta_{ij} \delta (\tau - \tau')
\eqno (2.1b)
$$
(the average denoted by an overline is with respect to the random force). In
particular this means that the correlation functions can be calculated using
either (1.1) or (2.1). 

In the Fokker-Planck formalism the correlation functions can be specified
in terms of the Green function $G(x_1^{(0)},\dots,x_N^{(0)}|
x_1,\dots,x_N;\tau)$, which is by definition the solution of (1.1)
subject to the initial condition
$$
p(x_1,\dots,x_N;\tau) \Big |_{\tau = 0} = \prod_{j=1}^N \delta (x_j - x_j^{(0)}).
\eqno (2.2)
$$
Thus for observables $A_\tau$ and $B_\tau$ (e.g.~$A_\tau = n_\tau (x) :=
\sum_{j=1}^N \delta (x - x_j(\tau))$, which corresponds to the
microscopic density) and initial p.d.f.~for the position of the particles $f$,
the
correlation between
$A_{\tau_a}$ and 
$B_{\tau_b}$ is defined as
$$
\langle  A_{\tau_a} B_{\tau_b} \rangle^T =
\langle  A_{\tau_a} B_{\tau_b} \rangle - \langle  A_{\tau_a}\rangle
\langle  B_{\tau_b} \rangle
\eqno (2.3a)
$$
where
\renewcommand{\theequation}{2.3b}
\begin{eqnarray}
\langle  A_{\tau_a} B_{\tau_b} \rangle & = &
\int_I dx_1^{(0)} \dots \int_I dx_N^{(0)} \, f(x_1^{(0)},\dots,x_N^{(0)})
\nonumber \\&& \times
\int_I dx_1^{(1)} \dots \int_I  dx_N^{(1)} \, A(\{ x_j^{(1)} \})
G(x_1^{(0)},\dots,x_N^{(0)}|x_1^{(1)},\dots,x_N^{(1)};\tau_a) \nonumber \\
& & \times \int_I dx_1^{(2)} \dots \int_I  dx_N^{(2)} \,
B(\{ x_j^{(2)} \}) G(x_1^{(1)},\dots,x_N^{(1)}| x_1^{(2)},\dots,x_N^{(2)};
\tau_b - \tau_a) \nonumber \\
\end{eqnarray}
\renewcommand{\theequation}{2.3c}
\begin{eqnarray}
 \langle  A_{\tau_a}\rangle & = &
\int_I dx_1^{(0)} \dots \int_I dx_N^{(0)} \, f(x_1^{(0)},\dots,x_N^{(0)})
\int_I dx_1 \dots \int_I  dx_N \, A(\{ x_j \}) \nonumber \\&&
G(x_1^{(0)},\dots,x_N^{(0)}|x_1,\dots,x_N;\tau_a),
\end{eqnarray}
and similarly the definition of $\langle  B_{\tau_b} \rangle$.
 The quantity
$\langle  A_{\tau_a} B_{\tau_b} \rangle$ is referred to as the distribution
function.

In the Langevin equation formalism the correlation between $A_\tau$ at time
$\tau = 0$ and $B_\tau$ at time $\tau_b$ is defined as
$$
\overline {\langle  A_0  B_{\tau_b} \rangle}^T_0 =
\overline {\langle  A_0  B_{\tau_b} \rangle}_0 - 
\overline {\langle  A_0 \rangle}_0
\overline {\langle  B_{\tau_b}  \rangle}_0
\eqno (2.4)
$$
where $\langle \:  \rangle_0$ denotes the average over the initial particle
distribution, while the overline denotes the average with respect to
the random force for times from 0 to $\tau$. The equivalence between
the Langevin and Fokker-Planck formalisms implies
$$
\langle  A_0  B_{\tau_b} \rangle^T = \overline {\langle  A_0  B_{\tau_b}
\rangle}^T_0
\eqno (2.5)
$$

When using the Fokker-Planck formalism we will make use of the general fact [11]
that
\renewcommand{\theequation}{2.6a}
\begin{eqnarray}
 e^{ \beta W /2} {\cal L}  e^{- \beta W /2} & = &
\sum_{j=1}^N \bigg ({1 \over \beta}{\partial^2 \over \partial \lambda_j^2}
-{\beta \over 4} \Big ( {\partial W \over \partial \lambda_j }\Big )^2
+ {1 \over 2} {\partial^2 W \over \partial \lambda_j^2} \bigg ) \nonumber \\
& = &
- {1 \over \beta} \sum_{j=1}^N \Pi^\dagger_j \Pi_j
\end{eqnarray}
where
$$
\Pi_j := {1 \over i} {\partial \over \partial x_j} - {i \beta \over 2}
{\partial W \over \partial x_j}.
\eqno (2.6b)
$$
Note that
$$
\Pi_j e^{-\beta W /2} = 0.
\eqno (2.7)
$$
The formula (2.6) shows that after conjugation with $ e^{-\beta W /2}$
the Fokker-Planck operator transforms into a Hermitian operator. For some
particular $W$, which include (1.2), we have
$$
 e^{ \beta W /2} {\cal L}  e^{- \beta W /2} = -{1 \over \beta}
(H - E_0)
\eqno (2.8)
$$
where $H$ is the Schr\"odinger operator for a quantum mechanical system with
one and two body interactions only, and $E_0$ is the corresponding ground state
energy. For example, with $W$ given by (1.2a) the equation (2.8) holds with
$$
H = - \sum_{j=1}^N {\partial^2 \over \partial x_j^2}
+{\beta^2 \over 4} \sum_{j=1}^N x_j^2 +\beta (\beta /2 - 1) \sum_{1 \le j < k \le N}
{1 \over (x_j - x_k)^2}.
\eqno (2.9)
$$
Note from (2.6a) and (2.8) that the ground state is (up to normalization)
given by
$$
\psi_0 = e^{-\beta W / 2}.
\eqno (2.10)
$$

\subsection{The current-current correlation}
In the Langevin description of Brownian dynamics the current-current
correlation is defined by choosing the observables $A_\tau$ and
$B_\tau$ in (2.4) as the classical current:
$$
A_0 = j_0(x_a), \quad B_\tau = j_\tau(x_b), \qquad
j_\tau(x) := \sum_{j=1}^N {d x_j(\tau) \over d \tau}
\delta (x - x_j(\tau)).
\eqno (2.11)
$$
In the Fokker-Planck description the classical current has no immediate
meaning as the velocities do not explicitly occur in (1.1). To define
the current in this situation we make use of the formula (2.6a) and
rewrite (2.3) in terms of time-dependent operators.

For this purpose we note from  (2.2) that
$$
G(x_1^{(0)},\dots,x_N^{(0)}|x_1^{(1)},\dots,x_N^{(1)};\tau) =
e^{{\cal L} \tau / \gamma} \prod_{l=1}^N \delta (x_l^{(1)} - x_l^{(0)})
\eqno (2.12)
$$
where the Fokker-Planck operator acts on $\{ x_l^{(1)} \}$ only.
Substituting (2.12) in (2.3b) allows the integration over $\{x_l^{(0)}\}$
to be carried out. Then substituting (2.12), with $\{x_l^{(0)}\}$,
$\{x_l^{(1)}\}$ replaced by $\{x_l^{(1)}\}$, $\{x_l^{(2)}\}$, in the result
and integrating over  $\{x_l^{(1)}\}$ we obtain
$$
\langle  A_{\tau_a} B_{\tau_b} \rangle =
 \int_I dx_1^{(2)} \dots \int_I  dx_N^{(2)} \,
 B_{\tau_b} e^{{\cal L} (\tau_b - \tau_a) / \gamma}
 A_{\tau_a} e^{{\cal L} \tau_a / \gamma}f(x_1^{(2)},\dots,x_N^{(2)}).
\eqno (2.13)
$$
Now it follows from (2.6a) that
$$
 e^{{\cal L} \tau / \gamma} = e^{-\beta W / 2}
e^{-\tau \sum_{j=1}^N \Pi^\dagger_j \Pi_j /\beta \gamma} e^{\beta W/2},
\eqno (2.14)
$$
so we can rewrite (2.13) as
$$
\langle  A_{\tau_a} B_{\tau_b} \rangle = 
 \int_I dx_1^{(2)} \dots \int_I  dx_N^{(2)} \, e^{-\beta W / 2}
B(\tau_b) A(\tau_a) f  e^{\beta W/2}
\eqno (2.15a)
$$
where
$$
 A(\tau) := e^{\tau \sum_{j=1}^N \Pi^\dagger_j \Pi_j /\beta \gamma}
A_{\tau}  e^{-\tau \sum_{j=1}^N \Pi^\dagger_j \Pi_j /\beta \gamma}
\eqno (2.15b)
$$
and similarly the definition of $B(\tau)$ (in deriving (2.15a) we have used the
fact that $ e^{\tau \sum_{j=1}^N \Pi^\dagger_j \Pi_j /\beta \gamma}
 e^{-\beta W / 2} = 1$, which follows from (2.7)).

Using (2.15b) we can define the current $j_\tau(x)$ by the
continuity equation
$$
{\partial \over \partial \tau} n(x,\tau) = - {\partial \over \partial
x} j(x,\tau),
\eqno (2.16)
$$
where $ n(x,\tau)$ is defined by the r.h.s.~of (2.15b) with 
$$
A_\tau = n_\tau(x) := \sum_{j=1}^N \delta (x - x_j(\tau))
\eqno (2.17)
$$
which is the microscopic density. Now (2.15b) is the imaginary time
quantum mechanical formula for the evolution of the operator
$A_\tau$ in the Heisenberg picture. Further the Hamiltonian
$\sum_{j=1}^N \Pi^\dagger_j \Pi_j$ is of the form
$$
-  \sum_{j=1}^N {\partial^2 \over \partial x_j^2} + V(x_1,\dots,x_N),
\eqno (2.18)
$$
so we conclude from the usual quantum mechanical verification of the
continuity equation that the sought definition of the current is
$$
j_\tau(x) = - {i\over \gamma \beta} 
\sum_{j=1}^N \Big ( {1 \over i} {\partial \over \partial x_j}
\delta (x - x_j(\tau)) + \delta (x - x_j(\tau))
{1 \over i} {\partial \over \partial x_j} \Big ).
\eqno (2.19)
$$
As written (2.19) assumes the particles are on a line. However the same formula
applies for each component of the current vector in higher-dimensions if we
simply replace $j_\tau(x)$ by $j_\tau^{(\alpha)}(x)$ and $\partial / 
 \partial x_j$ by $\partial /  \partial x_j^{(\alpha)}$.

Note that with $f$ proportional to $e^{-\beta W}$ (the equilibrium state)
in (2.15a) we can write
$$
\langle  A_{\tau_a} B_{\tau_b} \rangle = 
\langle \psi_0 | B(\tau_b) A(\tau_a) | \psi_0 \rangle
\eqno (2.20)
$$
which is precisely the quantum mechanical formula (with $\tau = it\beta \gamma$)
for ground state correlations. This shows that, up to a factor of
$(i/\gamma \beta)^2$, the current-current distribution for Brownian motion
described by the Fokker-Planck equation (2.1) with initial p.d.f.~given by the
equilibrium distribution is identical to the current-current distribution
of the corresponding quantum mechanical system (recall (2.8)). The
equivalence between the density-density distributions in this situation
has previously been shown by Beenakker and Rejaei [1].

\subsection{$f$-sum rule}
Denote the density-density correlation by $S$ so that
$$
S((x_a,0),(x_b,\tau_b)) = \Big \langle \sum_{j=1}^N \delta (x_a - x_j(0))
\sum_{k=1}^N \delta (x_b - x_k(\tau_b)) \Big \rangle^T
\eqno (2.21)
$$
where the average on the r.h.s.~is defined according to (2.3). When the initial
distribution $f$ in (2.2) is proportional to the equilibrium distribution
$e^{-\beta W}$ we have just remarked that $S$ is identical (with
$\tau = it\beta \gamma$) to the ground state density-density correlation
of the quantum system with Hamiltonian $\sum_{j=1}^N \Pi^\dagger_j
\Pi_j$. Since this Hamiltonian is of the form (2.18) it follows that
$S$ must satisfy the so called $f$-sum rule [10]:
$$
{\partial \over \partial \tau} \tilde{S}(k,\tau) \bigg |_{\tau = 0}
= - {k^2 \over \gamma \beta} \rho
\eqno (2.22a)
$$
where
$$
 \tilde{S}(k,\tau) := \int_{-\infty}^\infty S((x_a,0),(x_b,\tau))
e^{ik(x_a - x_b)} \, d(x_a - x_b)
\eqno (2.22b)
$$
and it is assumed that the ground state is translationally invariant so
that $S$ depends on $x_a - x_b$.  Subject to this assumption
(2.22a) is valid for $N$ finite as well as in the thermodynamic limit.

In fact it is possible to derive and indeed generalize the $f$-sum rule
entirely within the Brownian motion setting using either the Fokker-Planck or
Langevin equation formalism (for definiteness we will consider the latter). The
generalization is that we can take the initial distribution as proportional to
the Boltzmann factor
$e^{-\beta'W}$ (where we may have $\beta' \ne \beta$) and we do not assume
translational invariance.

We consider the quantity
$$
 \tilde{S}(x,k,\tau) := \int_{-\infty}^\infty S((x,0),(x_b,\tau))
e^{ikx_b} \, dx_b,
\eqno (2.23)
$$ 
where $S$ is given by (2.20) with the average now computed according to
the r.h.s.~of (2.5) because we will use the Langevin formalism. Differentiating
with respect to $\tau$ gives
$$
\gamma {\partial  \tilde{S}(x,k,\tau) \over \partial \tau} = ik
\overline{\Big \langle \sum_{j=1}^N \delta (x - x_j(0))
\sum_{l=1}^N  \gamma \dot{x}_l(\tau) e^{ikx_l(\tau)} \Big \rangle}_0.
\eqno (2.24)
$$
As $\tau$ approaches zero, we see from the Langevin equation (2.1) that
$$
\gamma \dot{x}_j(\tau) = -{\partial W \over \partial x_j} \bigg |_{\tau = 0}
+ {\cal F}_j(\tau) + O(\tau) 
\eqno (2.25a)
$$
and so
$$
 e^{ikx_l(\tau)} =  e^{ikx_l(0)}\Big ( 1 + {ik \over \gamma}
\int_0^\tau {\cal F}_l(\tau') \, d\tau' + O(\tau) \Big )
\eqno (2.25b)
$$
where the term proportional to $\tau$ not explicitly written is independent of
${\cal F}_j(\tau)$. Substituting (2.25) in (2.24) and using (2.1b) to
compute the average with respect to the random force gives
$$
\gamma {\partial  \tilde{S}(x,k,\tau) \over \partial \tau}
\Big |_{\tau = 0} = ik
\sum_{j,l=1}^N \Big \langle \delta (x - x_j(0))  e^{ikx_l(\tau)}
\Big ( - {\partial W \over \partial x_j} \bigg |_{\tau = 0}
+ {2 i k \over \beta} \int_0^\tau \delta (\tau - \tau') \, d\tau \Big )
\Big \rangle_0
\eqno (2.26)
$$

The delta function in (2.26) is non-zero on the boundary of the interval of
integration so we take one half of the value of the integrand. After averaging
over the initial distribution, the corresponding term of (2.26) is then
identified as proportional to $\tilde{S}(x,k,0)$. To simplify the term involving
$\partial W / \partial x_j$ in (2.26) we recall that the initial distribution
 is assumed proportional to $e^{-\beta' W}$. Since
$$
-{\partial W \over \partial x_j} e^{-\beta' W} = {1 \over \beta'}
{\partial  \over \partial x_j} e^{-\beta' W}
$$
integration by parts gives that this term can be rewritten as
$$
-{i k \over \beta'} S(x,k,0) + {1 \over \beta'}{\partial \over \partial x}
\Big ( e^{ikx} \rho (x) \Big ).
\eqno (2.27)
$$
Combining these results gives for our generalization of the $f$-sum rule
$$
 {\partial  \tilde{S}(x,k,\tau) \over \partial \tau}
\Big |_{\tau = 0} = -k^2 \Big ( {1 \over \beta \gamma} -
{1 \over \beta' \gamma} \Big ) \tilde{S}(x,k,0) + {ik \over \beta' \gamma}
{\partial \over \partial x} \Big ( e^{ikx} \rho (x) \Big ),
\eqno (2.28)
$$
which reduces to (2.22a) in the case $\beta' = \beta$, $\rho (x) = \rho$.

We emphasize that (2.28) holds for a finite system, as well as in the
thermodynamic limit. It has been presented in the one-dimensional case, but
by introducing the higher-dimensional Fourier transform in (2.23) and repeating
the working we see that (2.28) holds with $k^2$ replaced by $\sum_{\alpha}
(k^{(\alpha)})^2$ in the first term and $k$, $\partial / \partial x$, $kx$
replaced by $k^{(\alpha)}$, $\partial  / \partial x^{(\alpha)}$,
$k^{(\alpha)} x^{(\alpha)}$ (with summation over $\alpha$) in the last term.

\subsection{Static current-current distribution}
For the particular Schr\"odinger operator
$$
H = - \sum_{j=1}^N {\partial^2 \over \partial x_j^2}
 +\beta (\beta /2 - 1){\left ( \pi \over L \right )^2 } \sum_{1 \le j < k \le N}
{1 \over \sin^2 \pi (x_j - x_k)/L}, \quad 0 \le x_j \le L,
\eqno (2.29)
$$
which is related via equation (2.8) to the Fokker-Planck operator (1.1) with
potential (1.2b), it has been shown by Taniguchi et al.~[3] that the static
current-current distribution is given in terms of the static density-density
distribution by
$$
\langle \psi_0 | j(x_b,0) j(x_a,0) | \psi_0 \rangle
= - {\beta (\pi / L )^2 \over \sin^2 \pi (x_b - x_a) / L}
\langle \psi_0 | \rho (x_b,0) \rho (x_a,0) | \psi_0 \rangle, \quad x_a \ne x_b.
\eqno (2.30)
$$
From the remarks below (2.20), (2.30) is equivalent to the statement that for
the Fokker-Planck system with potential (1.2b) and initial p.d.f. equal
to the equilibrium distribution,
$$
\langle j_0(x_a) j_0(x_b) \rangle = -{\beta (\pi / L)^2 \over
\sin^2 \pi (x_b - x_a) /L} \Big ( {i \over \gamma \beta} \Big )^2
\langle n_0 (x_b) n_0(x_a) \rangle.
\eqno (2.31)
$$
The equation (2.30) was derived in ref.~[3] using the factorization (2.6a).
The same method readily extends to provide an analogous result for all 
Fokker-Planck systems in which $W$ consists of one and two body potentials
only.

Thus we are considering the average (2.15a) with 
$\tau_a$ and $\tau_b$ equal to zero, the observables
$A$ and $B$ given by the current (2.19) and the initial distribution $f$
proportional to $e^{-\beta W}$. To apply the method of ref.~[3], we
substitute for $\partial / \partial x_j$ in (2.19) using $\Pi_j$ as defined by
(2.6b) to obtain
$$
j_0(x) = -{i \over \beta \gamma} \sum_{j=1}^N \Big (
\Pi^\dagger_j \delta (x - x_j) + \delta (x - x_j) \Pi_j \Big )
\eqno (2.32)
$$
From this formula and (2.7) we have
$$
\langle j_0(x_a) j_0(x_b) \rangle =
\Big ( {i \over  \beta \gamma} \Big )^2 \sum_{j,k=1}^N 
{1 \over \hat{Z}} \int_I dx_1^{(2)} \dots \int_I  dx_N^{(2)} \, e^{-\beta W / 2}
\delta(x_b - x_j) \Pi_j \Pi_k^\dagger \delta(x_a - x_k)  e^{-\beta W/2}
\eqno (2.33)
$$
where
$$
 \hat{Z} :=  \int_I dx_1^{(2)} \dots \int_I  dx_N^{(2)} \, e^{-\beta W }.
$$
Let us now suppose $x_a \ne x_b$. Then the $j = k$ term vanishes. But for $j
\ne k$ we have
$$
[ \Pi_j^\dagger, \Pi_k^\dagger] = \beta {\partial^2 W \over \partial x_j
\partial x_k}.
\eqno (2.34)
$$
Using this formula, the fact that $\Pi_j \delta(x - x_k) = \delta (x - x_k)
\Pi_j$ for $j \ne k$, the property (2.7) and the definition (2.17) we deduce
from (2.33) that
$$
\langle j_0(x_a) j_0(x_b) \rangle = \Big ( {i \over  \beta \gamma} \Big )^2
\beta {\partial^2 W \over \partial x_1
\partial x_2} \bigg |_{x_1 = x_a \atop x_2 = x_b} \langle n_0(x_a) n_0(x_b)
\rangle,
\eqno (2.35)
$$
which is the sought formula.
We remark that it is also possible to derive (2.35) within the Langevin
equation setting used in the derivation of the $f$-sum rule given in the
previous section. Also, 
for $W$ given by (1.2b), note that (2.35) reduces to (2.31). 

Our derivation has been
presented in the one-dimensional case. The same result applies in higher
dimensions for the correlations between components $j^{(\alpha)}_0(x_a)$,
$j^{(\beta)}_0(x_b)$ of the current 
vector provided the $x_1,x_a$ and $x_2,x_b$ outside the average on the r.h.s.
are the components $(\alpha)$ and $(\beta)$ respectively of the 
corresponding position vector.

\subsection{Small-$k$ behaviour of structure factor}
In this section we will consider Brownian motion described by the Fokker-Planck
equation (1.1), with the potential $W$ consisting of general one and two body
terms
$$
W = \sum_{j=1}^N V_1(x_j) + \sum_{1 \le j < k \le N} V_2(|x_j - x_k|)
\eqno (2.36)
$$
such that the potential $V_2$ is non-integrable at infinity. For the particular
case of this type 
$$
V_2(x) = - \log |x|
\eqno (2.37)
$$
Dyson [5] introduced the hydrodynamic
equation
$$
\gamma j(x,\tau) = - \rho (x,\tau) {\partial \over \partial x}
\Big ( V_1(x) - \int_{-\infty}^\infty dx' \,  \rho (x',\tau)  V_2(|x - x'|)
\Big )
\eqno (2.38)
$$
(note that the r.h.s.~represents the force density)
to study the large-wavelength density fluctuations of the system. Here
$j(x,\tau)$ and $\rho (x,\tau)$ represent smoothed, continuum approximations
to the microscopic current and density.

In the case (2.37) it has been deduced in ref.~[1]  that
for a uniform initial state
$$
\rho (x',0) = \rho
\eqno (2.39)
$$
the Fourier transform of the density-density correlation defined by (2.22)
has the small-$k$ behaviour
$$
\tilde{S}(k,\tau) \: \sim \: \tilde{S}(k,0) e^{-k^2 \tau \rho \tilde{V}_2(k)/
\gamma}
\eqno (2.40)
$$
where
$$
 \tilde{V}_2(k) := \int_{-\infty}^\infty V_2(|x|) e^{ikx} \, dx.
\eqno (2.41)
$$
Inspection of the derivation given in [1] shows (2.40) to be a consequence
of (2.38) independent of the particular $V_2$, so its validity for
general $V_2$ is tied to the asymptotic validity of the hydrodynamic
approximation (2.38). We expect (2.38) to correctly describe the
large-wavelength density fluctuations whenever the potential $V_2(|x|)$
is not integrable at infinity (for integrable potentials the 
`electric field term' involving the integral on the r.h.s.~of (2.38) is not
present, rather the force density is due to a pressure gradient; see the 
next subsection). Now for potentials not integrable at infinity it is
expected [13] that 
$$
\tilde{S}(k,0)  \: \sim \: {1 \over \beta' \tilde{V}_2(k)}
\eqno (2.42)
$$
(here $\beta'$ is used as in the discussion of the generalized $f$-sum rule,
and thus may be different to
$\beta$ which occurs in (1.1)) and so (2.40) reads
$$
\tilde{S}(k,\tau) \: \sim \: {1 \over \beta' \tilde{V}_2(k)}
e^{-k^2 \tau \rho \tilde{V}_2(k)/
\gamma}.
\eqno (2.43)
$$
Note that this expression is consistent with the generalized
$f$-sum rule (2.28) with $\rho(x) = \rho$ (note that the first term in (2.28) is
of a lower order than the second since, as is seen from (2.42),
 $\tilde{S}(k,0) \to 0$ as $k
\to 0$ whenever
$V_2(|x|)$ is not integrable at infinity). It is expected to be valid for
non-integrable potentials in higher dimensions provided (2.41) is suitably
modified.

One consequence of (2.43) is the sum rule
$$
\int_0^\infty S(k,\tau) \, d\tau \: \sim \: { \gamma \over \rho \beta' k^2
(\tilde{V}_2(k) )^2} \quad \mbox{as} \quad k \to 0.
\eqno (2.44)
$$
In the particular case when $V_2$ is given by (2.37) and $\beta' = \beta$,
we noted in Section 2.1 that $S(k,\tau)$ in the Fokker-Planck system 
is identical to
$S(k,\tau)$ for the ground state of the Calogero-Sutherland quantum system with 
Schr\"odinger operator (2.9). Now for a quantum fluid the integral (2.44) is
related to the compressiblity of the ground state. This is consistent
with the fact that for $V_2$  given by (2.37)
$$
\tilde{V}_2(k) \: \sim \: {\pi \over |k|}
\eqno (2.45)
$$
and so the r.h.s.~of (2.44) is a constant.

Another consequence of (2.43) is the evaluation of the large-$\tau$ mean square
displacement of a particle from its initial position. Thus from ref.~[14],
assuming this displacement diverges, $\tilde{S}(k,0)$ diverges at the origin and
that the state is homogeneous, we have for large-$\tau$
$$
\langle (x(\tau) - x(0))^2 \rangle \: \sim \: {2 \over \pi \rho^2} 
\int_0^{\rho c} {dk \over k^2} \Big ( \tilde{S}(k,0) - \tilde{S}(k,\tau) \Big )
\eqno (2.46)
$$
where $c \ll 1$ is some constant. To make further progress, suppose $V_2(r)
\sim r^{-\alpha}$, $0 < \alpha < 1$ as $r \to \infty$, so that [15]
$$
\tilde{V}_2(k) \: \sim \: { \pi |k|^{\alpha - 1} \over \Gamma (\alpha) 
\cos \pi \alpha / 2} \qquad {\rm as} \quad k \to 0.
\eqno (2.47)
$$
The bounds on $\alpha$ are required for the validity of (2.46). Indeed the bound
$\alpha < 1$ is required so that $ \tilde{S}(k,0)$ diverges at the origin, while
substituting (2.47) in (2.43) and then substituting the result in (2.46) and
 shows that for (2.46) to diverge as $\tau \to \infty$ we must have
$\alpha > 0$. To calculate the large-$\tau$ behaviour we make the substitutions
of the previous sentence and change variables $y = k^{\alpha + 1} \Big (\pi / 
\Gamma ( \alpha) \cos \pi \alpha / 2 \Big ) \tau \rho / \gamma$. This gives
$$
\langle (x(\tau) - x(0))^2 \rangle \: \sim \: c(\beta', \gamma, \rho)
\tau^{\alpha/(\alpha + 1)} \qquad {\rm as} \quad \tau \to \infty
\eqno (2.48a)
$$
where
$$
c(\beta', \gamma, \rho) = {2 \over \beta' \pi \rho^2}
{(\rho / \gamma )^{\alpha / (\alpha + 1)} \over \alpha + 1}
\Big ( {\Gamma(\alpha) \cos \pi \alpha / 2 \over \pi} \Big )^{1/(\alpha + 1)}
\int_0^\infty y^{-2 + 1/(1+\alpha)} (1-e^{-y}) \, dy,
\eqno (2.48b)
$$
which is the sought formula. In the case $\alpha > 1$ (integrable potential)
it is shown in ref.~[14] that $\langle (x(\tau) - x(0))^2 \rangle $ is
proportional to $\tau^{1/2}$, independent of $\alpha$.

For a one-dimensional translationally invariant initial state which is 
the equilibrium state, the small-$k$ behaviour (2.43) also implies the small-$k$
behaviour of the Fourier transformed current-current correlation
$$
\tilde{C}(k,\tau) := \int_{-\infty}^\infty \langle j_0(0) j_\tau(x)
\rangle^T e^{ikx} \, dx.
$$
To see this, we recall [1] that for a one-dimensional translationally invariant
system,
a consequence of the continuity equation (2.16) is that
$$
\tilde{C}(k,\tau) = {1 \over k^2} {\partial^2 \over \partial \tau_a
\partial \tau_b}
\tilde{S}(k;\tau_a,\tau_b)
\eqno (2.49)
$$
where
$$
\tilde{S}(k;\tau_a,\tau_b) := \int_{-\infty}^\infty
S((0,\tau_a),(x,\tau_b)) e^{ikx} \, dx.
$$
But by the assumption that the initial state is the equilibrium state
$\tilde{S}(k;\tau_a,\tau_b) = \tilde{S}(k,\tau_b-\tau_a)$ and so (2.49)
reads
$$
\tilde{C}(k,\tau) = -{1 \over k^2}{d^2 \over d \tau^2} \tilde{S}(k,\tau ).
\eqno (2.50)
$$
Substituting (2.50) in (2.43) (with $\beta' = \beta$) gives for the small-$k$
asymptotic formula
$$
\tilde{C}(k,\tau) \: \sim \: - { \rho^2 \over \beta \gamma^2 } k^2
\tilde{V}_2(k) e^{-k^2 \tau \rho \tilde{V}_2(k)/
\gamma}.
\eqno (2.51)
$$

For the Dyson log-gas, when $\tilde{V}_2(k)$ is given by (2.45), (2.51) gives
the $\tau = 0$ result
$$
\tilde{C}(k,0)  \: \sim \: - {\rho^2 \pi |k| \over \beta \gamma^2}.
$$
By taking the inverse transform we deduce the large-$x_{ab}$ behaviour
$$
\langle j_0(x_a) j_0(x_b) \rangle  \: \sim \: {\rho^2 \over \beta \gamma^2
(x_a - x_b)^2},
\eqno (2.52)
$$
which is consistent with the large-$x_{ab}$ behaviour of the thermodynamic
form of (2.31) , since in this limit $\langle n_0 (x_b) n_0(x_a) \rangle
\sim \rho^2$.

We also have the analogue of (2.44) in the setting of the validity of
(2.50). Thus integrating (2.50) with respect to $\tau$ gives
$$
\int_0^\infty  \tilde{C}(k,\tau) \, d\tau = -{1 \over k^2}
{d \over d \tau} \tilde{S}(k,\tau) \Big |_{\tau = 0} = {\rho \over 
\gamma \beta},
\eqno (2.53)
$$
where to obtain the last equality the $f$-sum rule (2.22a) has been used.
Note that unlike (2.44), the equation (2.53) is valid for all values of $k$.

\subsection{Second order correction}
To obtain the next order correction in the small-$k$ expansion (2.40) we
refine the hydrodynamic equation (2.38) to include a pressure gradient
$-\partial p(x,\tau) / \partial x$ as an additional force density on the
r.h.s..  The strategy for the solution of this equation is the same as given in
[1] for the solution of (2.40). The first step is to differentiate both sides of
the equation with respect $x$ and to then substitute for $\partial j(x,\tau) /
\partial x$ using the continuity equation (2.16). After linearizing in
$\delta \rho (x,\tau) := \rho (x,\tau) - \rho$ (note that the pressure is
linearized by  $p(x,\tau) - p = (\partial p / \partial \rho) \delta \rho
(x,\tau)$ where $p$ denotes the bulk pressure) and assuming the equilibrium
condition that for large-$L$
$$
{\partial \over \partial x} \Big ( V_1(x) - \rho \int_{-L}^L
 V_2(|x - x'|) \, dx' \Big )   = 0
$$
we obtain
$$
{\partial \delta \rho (x,\tau) \over \partial \tau} =
{\rho \over \gamma} {\partial^2 \over \partial x^2} \int_{-\infty}^\infty
 V_2(|x - x'|)  \delta \rho (x',\tau) \, dx' + {1 \over \gamma}
 {\partial p \over \partial \rho}{\partial^2 \over \partial x^2}  \delta \rho
(x,\tau)
$$
Taking the Fourier transform of both sides and integrating by parts twice
gives a simple differential equation for $ \delta \tilde{\rho} (k,\tau)$ which
has solution
$$
 \delta \tilde{\rho} (k,\tau) =  \delta \hat{\rho} (k,0)
\exp \bigg ( - \Big ( k^2 \rho \tilde{V}_2(k) + {\partial  p \over \partial \rho}
k^2 \Big ) {\tau \over \gamma} \bigg ).
\eqno (2.54)
$$
But
$$
\tilde{S}(k,\tau)  = {1 \over L} \langle \delta  \tilde{\rho} (-k,0) \delta 
\tilde{\rho}
(k,\tau)
\rangle.
\eqno (2.55)
$$
Substituting (2.54) gives the desired second order correction to (2.40):
$$
\tilde{S}(k,\tau) \: \sim \: \tilde{S}(k,0)
\exp \bigg ( - \Big ( k^2 \rho \tilde{V}_2(k) + {\partial  p \over \partial \rho}
k^2 \Big ) {\tau \over \gamma} \bigg ).
\eqno (2.56)
$$
Note that for consistency with the second term of the $f$-sum rule (2.28) 
with $\rho (x) = \rho$, which gives the leading small-$k$ behaviour, we
must have
$$
\tilde{S}(k,0) \: \sim \: {1 \over \beta'\Big ( \tilde{V}_2(k) + {1
\over \rho}{\partial  p
\over \partial \rho} \Big )}.
\eqno (2.57)
$$
A thermodynamic derivation of this result is given in the appendix.

\section{Exact results}
\subsection{GOE $\to$ GUE transition}
Consider the Dyson Brownian motion model specified by the Fokker-Planck
equation (1.1) with $\beta = 2$, $W$ given by (1.2a) and the initial
distribution proportional to $e^{- W}$. Since $e^{- W}$ is
proportional to the eigenvalue p.d.f.~for the Gaussian Orthogonal Ensemble
(GOE) of  random real symmetric matrices, while $e^{-2 W}$ is proportional
to the eigenvalue p.d.f.~for the Gaussian Unitary Ensemble (GUE) of random
Hermitian matrices, this specific Dyson Brownian motion model describes the
GOE $\to$ GUE transition. In the $N \to \infty$ limit (suitably scaled)
the system is translationally invariant with bulk density $\rho$ and the
Fourier transform of the density-density correlation (2.22) is given exactly by
[4, eq.(3.32) with the correction that an additional factor 
$e^{(\tau \pi |k| \rho /\gamma) (1 - |k|/2 \pi \rho)}$ be included in the first
term of (3.32c)]
\renewcommand{\theequation}{3.1a}
\begin{eqnarray}
\tilde{S}(k,\tau) & = & {2 \gamma \over \pi \tau |k|} \exp(-\tau \pi |k|
\rho /\gamma) \mbox{sinh}(\tau k^2 /2\gamma) \nonumber \\
&  & - {|k| \over 2 \pi} e^{\tau k^2 /2\gamma} \int_1^{1 + |k|/\pi \rho}
dk_1 {1 \over k_1} e^{- \pi \tau \rho |k| k_1 /\gamma}, \qquad
0 \le |k| \le 2 \pi \rho
\end{eqnarray}
\renewcommand{\theequation}{3.1b}
\begin{eqnarray}
\tilde{S}(k,\tau) & = & {2 \gamma \over \pi \tau |k|} \exp(-\tau k^2 /2\gamma)
\mbox{sinh}(\tau \pi |k|\rho /\gamma) \nonumber \\
&  & - {|k| \over 2 \pi} e^{\tau k^2 /2\gamma} \int_{-1 + |k|/\pi \rho}^{1 +
|k|/\pi
\rho} dk_1 {1 \over k_1} e^{- \pi \tau \rho |k| k_1 /\gamma}, \qquad
|k| \ge 2 \pi \rho
\end{eqnarray}
This result provides an illustration of the generalized $f$-sum rule (2.20)
in the case $\beta \ne \beta'$. Thus according to (2.28) with $\rho(x) =
\rho$, $\beta = 2$, $\beta'=1$ and $x=0$, for all $k$ (3.1) must satisfy
$$
 {\partial  \tilde{S}(k,\tau) \over \partial \tau}
\bigg |_{\tau = 0} =  { k^2 \over 2 \gamma} 
 \tilde{S}(k,0) - {k^2 \rho \over  \gamma}
\eqno (3.2)
$$
An elementary calculation verifies that this is indeed so.
Also, the small-$k$ expansion of (3.1a) is consistent with (2.56) and
(2.57) with $\beta'=1$, $\tilde{V}_2(k)$ given by (2.45) and $\partial p
/ \partial \rho$ calculated from the equation of state
$\beta' p = (1 - \beta'/2) \rho$.

\subsection{Free fermion type system}
As remarked earlier, for a number of special choices of $W$ including (1.2),
the equation (2.8) holds relating the Fokker-Planck operator to a Schr\"odinger
operator. A further property of this Schr\"odinger operator, as is illustrated
by (2.9), is that at the special coupling $\beta = 2$ the coefficient of the
two-body term vanishes and the system can be regarded as free fermions in
an external potential.

Consider such a situation in which the one-particle Schr\"odinger operator,
$H_1$ say, has a complete set of real, orthonormal eigenfunctions 
$\{\psi_k\}_{k=0,1,\dots}$ and corresponding eigenvalues $\{\epsilon_k +
E_0 \}_{k=0,1,\dots}$ and the particles are confined to an interval $I$.
A standard calculation (see e.g.~ref.[1]) gives that the density-density
correlation for the $N$-particle system is given by
\renewcommand{\theequation}{3.2}
\begin{eqnarray}
S((x,0),(x',\tau)) & = & \sum_{p=0}^\infty \psi_p(x)  \psi_p(x') 
e^{-\epsilon_p \tau / \gamma \beta} \sum_{q=0}^{N-1}  \psi_q(x)  \psi_{q}(x')
e^{\epsilon_q \tau/ \gamma \beta} \nonumber \\
& & - \sum_{p=0}^{N-1}  \psi_p(x)  \psi_p(x') e^{-\epsilon_p \tau / \gamma
\beta} \sum_{q=0}^{N-1}  \psi_q(x)  \psi_{q}(x')
e^{\epsilon_q \tau/ \gamma \beta}
\end{eqnarray}
Since here we have set $t = \tau / i\beta \gamma$, this represents the
density-density correlation for the corresponding Fokker-Planck system with
initial state equal to the equilibrium state. This exact result can be used to
illustrate the generalized $f$-sum rule (2.20) in the case of finite $N$ and
non-uniform density $\rho (x)$. 

Thus we want to show that (3.2) satisfies
$$
 {\partial  \tilde{S}(x,k,\tau) \over \partial \tau}
\Big |_{\tau = 0} = {ik \over \beta \gamma}
{\partial \over \partial x} \Big ( e^{ikx} \rho (x) \Big ),
\eqno (3.3)
$$
where $ \tilde{S}(x,k,\tau)$ is defined by (2.23) with $S$ therein given by 
(3.2), and $\rho (x)$ is given by
$$
\rho (x) = \sum_{p=0}^{N-1} \Big ( \psi_p(x) \Big )^2
\eqno (3.4)
$$
(this latter formula follows from the free fermion calculation). For this
purpose we note that differentiating the second term in (3.2) with respect to
$\tau$ and then setting $\tau = 0$ gives zero as the two terms which result
cancel. Using this feature and the fact $(H_1 - E_0) \psi_q(x') =
\epsilon_q \psi_q(x')$ we see that
\renewcommand{\theequation}{3.5}
\begin{eqnarray}
 {\partial \over  \partial \tau} S((x,0),(x',\tau)) \Big |_{\tau = 0}& = &
- {1 \over \gamma \beta}  \sum_{p=0}^\infty \psi_p(x) \Big ( (H_1 -
E_0)\psi_p(x') \Big )  \sum_{q=0}^{N-1} 
\psi_q(x) 
\psi_{q}(x') \nonumber \\
& & + {1 \over \gamma \beta} \sum_{p=0}^{\infty}  \psi_p(x)  \psi_p(x')
 \sum_{q=0}^{N-1}  \psi_q(x)  (H_1 - E_0)\psi_{q}(x').
\end{eqnarray}
Taking the Fourier transform of both sides with respect to $x'$ and integrating
 by parts (twice) the first term on the r.h.s.~using the
explicit formula
$$
H_1 = -{d^2 \over d x'^2} + V(x')
$$
we see that the last term cancels. Furthermore, noting by completeness that
$$
\sum_{p=0}^\infty \psi_p(x') \psi_p(x) = \delta (x' - x)
$$
we obtain
\renewcommand{\theequation}{3.6}
\begin{eqnarray}
 {\partial \over  \partial \tau} S((x,0),(x',\tau)) \Big |_{\tau = 0}& = &
{2 i k \over \gamma \beta} \int e^{i k x'} \delta (x - x')
 {\partial \over \partial x'} \sum_{q=0}^{N-1}  \psi_q(x)  \psi_q(x')
\, dx'  \nonumber \\
&& - {k^2 \over \gamma \beta} \int  e^{i k x'} \delta (x - x')
\sum_{q=0}^{N-1}  \psi_q(x)  \psi_q(x')
\, dx'  \nonumber \\
& = & { i k \over \gamma \beta} e^{ikx} {d \over dx} \rho(x) - {k^2 \over \gamma
\beta}  e^{ikx}  \rho(x)
\end{eqnarray}
where to obtain the last equality (3.4) has been used.
Comparing (3.6) with the $f$-sum rule formula (3.3) shows that the required
agreement with that result has been demonstrated.

\subsection{The Dyson Brownian motion model for rational $\beta$}
In the thermodynamic limit the density-density correlation $S(x,\tau)$ for the
Dyson  Brownian motion model specified by the Fokker-Planck equation (1.1)
with $W$ given by (1.2b) has been calculated for all rational values of
$\beta$. With $\beta / 2 := \lambda = p/q$ ($p$ and $q$ relatively prime) the
result is [9]
\renewcommand{\theequation}{3.7a}
\begin{eqnarray}\lefteqn{
\lim_{N,L \rightarrow \infty \atop N/L = \rho}{S}(x,\tau)}
\nonumber  \\& &
=  C_{p,q}(\lambda)
  \prod_{i=1}^{q}\int_0^\infty dx_i 
 \prod_{j=1}^p \int_0^1 dy_j Q_{p,q}^2
 F(q,p,\lambda|\{x_i,y_j\})\,
 \cos Q_{p,q}x \,\exp(-E_{p,q}\tau/2 \lambda \gamma) \nonumber \\
\end{eqnarray}
where the momentum $Q$ and the energy $E$ variables are given by
$$
Q_{p,q} := 2 \pi \rho \Big ( \sum_{i=1}^q x_i + \sum_{j=1}^p y_j \Big ), 
\quad
E_{p,q} := (2 \pi \rho )^2 \Big ( \sum_{i=1}^q \epsilon_P(x_j)
+ \sum_{j=1}^p \epsilon_H(y_j) \Big )
\eqno (3.7b)
$$
with
$$
\epsilon_P(x) = x(x+\lambda) \quad  \mbox{ and} \quad \epsilon_H(y) = \lambda y
(1-y),
\eqno (3.7c)
$$
the form factor $F$ is given by
$$
F(q,p,\lambda|\{x_i,y_j\}) = \prod_{i=1}^q \prod_{j=1}^p (x_i + \lambda y_j)^{-2}
{\prod_{i<i'}|x_i - x_{i'}|^{2 \lambda}\prod_{j<j'}|y_j - y_{j'}|^{2 /\lambda}
\over \prod_{i=1}^q (\epsilon_P(x_i))^{1-\lambda} \prod_{j=1}^p
(\epsilon_H(y_j))^{1 - 1/\lambda} }
\eqno (3.7d)
$$
and the normalization is given by
$$
C_{p,q}(\lambda ) = {\lambda^{2p(q-1)} \Gamma^2(p) \over 2 \pi^2 p! q!}
{\Gamma^q(\lambda) \Gamma^p(1/\lambda) \over 
\prod_{i=1}^q \Gamma^2(p-\lambda (i-1)) \prod_{j=1}^p \Gamma^2(1-(j-1)/\lambda)}.
\eqno (3.7e)
$$ 

This exact result can be used to illustrate the sum rules (2.43) and (2.57).
To begin, note that $S(x,\tau)$ is even in $x$ and so $\tilde{S}(k,\tau)$
is even in $k$. It therefore suffices to consider the case $k > 0$. With
this assumption, from (3.7)
\begin{eqnarray*}\lefteqn{
\tilde{S}(k,\tau)}\\ & & = \pi C_{p,q}(\lambda) \prod_{i=1}^q
\int_0^\infty dx_i 
 \prod_{j=1}^p \int_0^1 dy_j Q_{p,q}^2
 F(q,p,\lambda|\{x_i,y_j\})\,
 \delta(k - Q_{p,q}) \,\exp(-E_{p,q}\tau/2 \lambda \gamma).
\end{eqnarray*}
Changing variables $x_i \mapsto k x_i$, $y_j \mapsto ky_j$ gives
\renewcommand{\theequation}{3.8a}
\begin{eqnarray}\lefteqn{
\tilde{S}(k,\tau) = \pi k C_{p,q}(\lambda) e^{-k \pi \rho \tau / \gamma}}
\nonumber \\ & &
\times
\prod_{i=1}^q
\int_0^\infty dx_i 
 \prod_{j=1}^p \int_0^1 dy_j Q_{p,q}^2
 \hat{F}(q,p,\lambda|\{x_i,y_j\};k)\,
 \delta(1 - Q_{p,q}) \,\exp(-\hat{E}_{p,q,k}\tau/2 \lambda \gamma). \nonumber \\
\end{eqnarray}
where
$$
 \hat{F}(q,p,\lambda|\{x_i,y_j\};k) = \prod_{i=1}^q \prod_{j=1}^p (x_i
+ \lambda y_j)^{-2}
{\prod_{i<i'}|x_i - x_{i'}|^{2 \lambda}\prod_{j<j'}|y_j - y_{j'}|^{2 /\lambda}
\over \prod_{i=1}^q \Big ( x_i (kx_i + \lambda) \Big )^{1 - \lambda}
 \prod_{j=1}^p \Big ( \lambda y_j (1 - y_j k) \Big )^{1 - 1/\lambda}}
\eqno (3.8b)
$$
and
$$
\hat{E}_{p,q,k}= (2 \pi \rho)^2 k^2 \Big ( \sum_{i=1}^q x_i^2 - \lambda
\sum_{j=1}^p y_j^2 \Big ).
\eqno (3.8c)
$$

In the limit $k \to 0$ the integral in (3.8a) is independent of $k$, so to
leading order we have
$$
\tilde{S}(k,\tau) \: \sim \: \pi |k| C_{p,q}(\lambda) I(\lambda)  e^{-|k| \pi
\rho
\tau /
\gamma}
\eqno (3.9)
$$
where
$$
 I(\lambda) =  \prod_{i=1}^q
\int_0^\infty dx_i 
 \prod_{j=1}^p \int_0^\infty dy_j Q_{p,q}^2 G(q,p,\lambda|\{x_i,y_j\})
\delta(1-Q_{p,q})
\eqno (3.10) 
$$
with
$$
G(q,p,\lambda|\{x_i,y_j\}) = \prod_{i=1}^q \prod_{j=1}^p (x_i + \lambda y_j)^{-2}
{\prod_{i<i'}|x_i - x_{i'}|^{2 \lambda}\prod_{j<j'}|y_j - y_{j'}|^{2 /\lambda}
\over \prod_{i=1}^q  x_i^{1 - \lambda}
 \prod_{j=1}^p  \lambda y_j^{1 - 1/\lambda}}.
\eqno (3.11)
$$
To evaluate (3.10) we rewrite it as
$$
 I(\lambda) =  \lim_{\epsilon \to 0^+} {1 \over 2 \pi}
\int_{-\infty}^\infty du \, e^{iu} \, \prod_{i=1}^q
\int_0^\infty dx_i 
 \prod_{j=1}^p \int_0^\infty dy_j Q_{p,q}^2 G(q,p,\lambda|\{x_i,y_j\})
e^{-iQ_{p,q}u} e^{-\epsilon Q_{p,q}}.
\eqno (3.12)
$$ 
Now change variables $x_i \mapsto {1 \over \epsilon + iu} x_i$,
$y_j \mapsto {1 \over \epsilon + iu} y_j$. This shows 
$$
 I(\lambda) =
\Big (  \lim_{\epsilon \to 0^+} {1 \over 2 \pi}
\int_{-\infty}^\infty du \, {e^{iu} \over (\epsilon + iu)^2} \Big )
 \prod_{i=1}^q \int_0^\infty dx_i  \prod_{j=1}^p \int_0^\infty dy_j Q_{p,q}^2
G(q,p,\lambda|\{x_i,y_j\})
 e^{-\epsilon Q_{p,q}}.
\eqno (3.13)
$$
The first integral above equals unity in the limit $\epsilon \to 0^+$, while
the multiple integral has been evaluated in an earlier work
[16] as equal to $1/(C_{p,q}(\lambda) \pi^2 \beta)$. Substituting in (3.13)
and then substituting the result in (3.9) gives the hydrodynamic result
(2.43) with the substitution (2.45).

It is also possible to illustrate (2.57) by expanding 
 (3.8a) to the next order in $k$:
$$
 \hat{F}(q,p,\lambda|\{x_i,y_j\};k) \: \sim \: 
G(q,p,\lambda|\{x_i,y_j\})\Big ( 1 + k (1-1/\lambda)Q_{p,q}/2 \pi \rho \Big ).
$$
This gives 
\renewcommand{\theequation}{3.14}
\begin{eqnarray}\lefteqn{
\tilde{S}(k,0) \: \sim \: {|k| \over \pi \beta}} \nonumber \\&& + {1 \over 2 \pi
\rho} (1 - 1/\lambda) \pi k^2 C_{p,q}(\lambda) 
\prod_{i=1}^q
\int_0^\infty dx_i 
 \prod_{j=1}^p \int_0^\infty dy_j Q_{p,q}^3 G(q,p,\lambda|\{x_i,y_j\})
\delta(1-Q_{p,q}). \nonumber \\
\end{eqnarray}
Comparing the multiple integral above with that in (3.10) shows that they
are identical apart the factor of $ Q_{p,q}^3$ which reads $ Q_{p,q}^2$
in (3.10). But this factor does not effect the value of the integral due to
factor of $\delta(1-Q_{p,q})$. Thus from the above working we have
$$
\tilde{S}(k,0) \: \sim \: {|k| \over \pi \beta} + {1 \over 2 \pi \rho}
\Big ( {1 \over \pi \beta} \Big ) \Big ( { \beta - 2 \over \beta} \Big ) k^2.
\eqno (3.15)
$$
This is precisely the result (2.57) expanded to second order in $k$ with
$\beta' = \beta$, the
substitution (2.45) and the partial derivative evaluated from the equilibrium
equation of state 
$\beta p = (1 - \beta/2) \rho$.

\section{Conclusion}
We have presented a systematic study of exact and asymptotic formulas relating
to dynamical correlations for overdamped Brownian motion. Our main new results
are the operator formula (2.19) for the current in the Fokker-Planck
description, the generalized $f$-sum rule (2.28), the formula (2.35) relating
the static current-current correlation to the static density-density
correlation and the asymptotic formula (2.52) for the small-$k$
behaviour of the dynamical structure factor. We have illustrated these
formulas using  known exact evaluations of the dynamical density-density
in certain special models.

Our work builds on the results presented in refs.~[1-3]. In this regard we
point out that in ref.~[1] the current is defined not in terms of 
$dx_j(\tau)/d\tau$ as in (2.11), but rather with this factor replaced
by $dx_j(\tau)/d\tau^{1/2}$. The latter choice is appropriate in applications
of the Dyson Brownian motion model to chaotic spectra as the perturbing
parameter $X$ is related to $\tau$ by $\tau = X^2$.

\vspace{.5cm}
\noindent
{\Large \bf Acknowledgements}
\vspace{.2cm}

\noindent
The visit of PJF to Orsay was financed by the Bede Morris fellowship scheme
through his selection as  The 1996 French Embassy Fellow.

\pagebreak
\noindent
{\Large \bf Appendix}

\vspace{.2cm}
\noindent
Here we will derive (2.57) using a thermodynamic argument. For notational
convenience we will assume the domain is an interval of length $L$,
although our argument applies equally well in higher dimensions. The setting
is a uniform one-component system with a pair potential non-integrable at
infinity. Such a system requires a neutralizing background for thermodynamic
stability, and supposing the particles carry unit charge, the particle and
charge density fluctuations are identical. We consider the  microscopic
density and suppose it has a fluctuation of the form 
$$
\delta \rho (x) = \rho_k e^{ikx} + \rho^*_k e^{-ikx}.
\eqno ({\rm A}1)
$$
Here $\rho_k$ is $1/L$ times the Fourier transform of the
microscopic density.
If $2 \pi / |k|$ is a macroscopic wavelength, $\rho_k$ is a good collective
variable which can be used instead of the particle coordinates.
The system will gain an
electrostatic energy
$$
{1 \over 2} \int_{-L/2}^{L/2} dx' \, \delta \rho (x')
 \int_{-L/2}^{L/2} dx \,  \delta \rho (x) V_2(|x - x'|).
\eqno ({\rm A}2)
$$

Let $f(\rho)$ be the free energy per unit length of the locally neutral system.
To second order in $\delta \rho (x)$ the density fluctuation  changes this free
energy by an amount
$$
\delta f = {\partial f \over \partial \rho} \Big |_{\rho = \rho_0}
\delta \rho (x) + {1 \over 2} {\partial^2 f \over \partial \rho^2} 
\Big |_{\rho =
\rho_0}\Big (\delta \rho (x) \Big )^2. 
\eqno ({\rm A}3)
$$
Hence the total change in free energy due to the
fluctuation is
\renewcommand{\theequation}{A4}
\begin{eqnarray}
\delta F & = &  {1 \over 2} \int_{-L/2}^{L/2} dx' \, \delta \rho (x')
 \int_{-L/2}^{L/2} dx \,  \delta \rho (x) V_2(|x - x'|) +
{1 \over 2} {\partial^2 f \over \partial \rho^2} 
\Big |_{\rho =
\rho_0}  \int_{-L/2}^{L/2} dx \, \Big (\delta \rho (x) \Big )^2 \nonumber \\
& \sim & {L |\rho_k|^2 \over 2} \Big ( \tilde{V}_2(k) + 
{1 \over 2} {\partial^2 f \over \partial \rho^2} 
\Big |_{\rho =
\rho_0} \Big ).
\end{eqnarray}

In general the probability density function for an event in the system is
proportional to $e^{-\beta' \delta F}$ where $\delta F$ is the corresponding
change in the free energy. From  (2.55), where now $\delta \tilde{\rho} (k)
= L \rho_k$, we have
\renewcommand{\theequation}{A5}
\begin{eqnarray}
\tilde{S}(k)& = &L \langle |\rho_k|^2 \rangle. 
\end{eqnarray}
Computing the r.h.s.~of (A5) from the Gaussian distribution $e^{-\beta' \delta
F}$ gives (2.57).

\pagebreak

\begin{description}
\item[References]
\item[][1] C.W.J.~Beenakker and B. Rejaei, Physica A {\bf 203}, 61
(1994)
\item[][2] E.R.~Mucciolo,  B.S. Shastry , B.D. Simons and B.L. Altshuler, Phys.
Rev. B {\bf 49}, 15197 (1994)
\item[][3] N. Taniguchi, B.S. Shastry and B.L. Altshuler, Phys. Rev.
Lett, {\bf 75}, 3724 (1995)
\item[][4] P.J.~Forrester, Physica A {\bf 223}, 365 (1996)
\item[][5] F.J.~Dyson, J.~Math.~Phys.~{\bf 3},
1199 (1962)
\item[][6] F.~Haake, {\it Quantum Signatures of Chaos},
(Springer, Berlin, 1992)
\item[][7] A.M.S.~Mac\^{e}do, Phys.~Rev.~B {\bf 49}, 16841 (1994)
\item[][8] K.~Frahm and J.-L.~Pichard, J.~Phys. I (France) {\bf 5} , 877 (1995) 
\item[][9] Z.N.C.~Ha, Nucl.~Phys.~B {\bf 435}, 604 (1995)
\item[][10] D.~Pines and P.~Nozieres, {\it The Theory of Quantum Liquids},
(W.A.~Benjamin, New York, 1966)
\item[][11] N.G.~van Kampen, {\it Stochastic Processes in Physics and
Chemistry} (North-Holland, Amsterdam, 1981) 
\item[][12] H.~Risken, {\it The Fokker-Planck Equation},
(Springer, Berlin, 1992)
\item[][13] L.~Reatto and G.V.~Chester, Phys.~Rev. {\bf 155}, 88 (1967)
\item[][14] H.~Spohn, J.~Stat.~Phys. {\bf 47}, 669 (1987)
\item[][15] J.~Lighthill, {\it Introduction to Fourier Analysis and
Generalized Functions}, (CUP, Cambridge, 1958)
\item[][16] P.J.~Forrester and J.A.~Zuk, Nucl.~Phys.~B {\bf 473}, 616 (1996)
\end{description}

\end{document}